\documentclass[aps,prb,twocolumn,showpacs,floats,preprintnumbers,amsmath,amssymb]{revtex4}

\usepackage{graphicx}
\usepackage{dcolumn}
\usepackage{bm}

\begin{document}

\preprint{(submitted to Phys. Rev. B)}

\title{Azobenzene at Coinage Metal Surfaces:\\The Role of Dispersive van der Waals Interactions}

\author{Erik McNellis}
\author{J\"org Meyer}
\author{Karsten Reuter}
\affiliation{Fritz-Haber-Institut der Max-Planck-Gesellschaft, Faradayweg 4-6, D-14195 Berlin (Germany)}

\received{23 June 2009}

\begin{abstract}
We use different semi-empirical dispersion correction schemes to assess the role of long-range van der Waals interactions in the adsorption of the prototypical molecular switch azobenzene (C$_{6}$H$_{5}\!-\!$N$_{2}\!-\!$C$_{6}$H$_{5}$) at the coinage metal surfaces Cu(111), Ag(111) and Au(111). Compared to preceding density-functional theory results employing a semi-local exchange and correlation functional we obtain partly sizable changes of the computed adsorption geometry and energetics. The discomforting scatter in the results provided by the different schemes is largely attributed to the unknown form of the damping function in the semi-empirical correction expression. Using the congeneric problem of the adsorption of benzene as a vehicle to connection with experiment, we cautiously conclude that the account of dispersive interactions at the metal surfaces provided by the various schemes is in the right ballpark, with the more recent, general schemes likely to overbind.
\end{abstract}



\maketitle

\section{Introduction}

Since the advent of semiconductor based microelectronics, the component density of integrated circuits has increased at a steady exponential rate. As component sizes approach the nano-scale, maintaining this pace will require a shift towards alternative materials more versatile than doped silicon. One such approach, already on the research agenda for several years, is to exploit the great variety and adaptability of organic molecules in component design. In the context of such a molecular nanotechnology, molecules with properties bi-stably and reversibly modifiable by external stimuli, so-called molecular switches, are a research topic of paramount importance. The azobenzene molecule (C$_{6}$H$_{5}\!-\!$N$_{2}\!-\!$C$_{6}$H$_{5}$) qualifies in this class by undergoing a reversible trans-cis isomerization. Because of its relative chemical simplicity, this molecule has come under intensive experimental scrutiny. Numerous potential applications have been proposed, e.g. as a light-driven actuator\cite{yu03}, or data storage medium\cite{liu90}. On the theoretical side, the precise mechanism behind the switching from the planar, C$_{2h}$ symmetric trans isomer to the torsioned-twisted C$_{2}$ cis isomer, is a much debated topic. The consensus view explains the observed trans-cis (cis-trans) isomerization, following photo-excitation of the $\pi \rightarrow \pi^{*}$ ($n \rightarrow \pi^{*}$) resonance, in terms of conical intersections between the ground and low lying excited states on the isomerization reaction coordinate\cite{cembran04,diau04,ishikawa01}. 

Nevertheless, gas phase or solvated switches carry their own set of limitations within the framework of molecular microelectronics, wherefore in recent years, properties of switches adsorbed at solid surfaces\cite{applphysa} has emerged as an important research field. Since the switching function is an innate property of the azobenzene electronic structure, the choice of substrate and substrate coupling is non-trivial: While bonding strong enough to localize and order switches is desirable, significant hybridization of the mentioned frontier orbitals, or steric hindrance due to substrate registry is not. Close-packed coinage metal (Cu, Ag, Au) surfaces ostensibly offer a reasonable such balance. However, even at these substrates, azobenzene and -derivate switches exhibit a host of modifications to the switching function, in all combinations of reversible or irreversible switching, and switching by light or exclusively by a local electric field\cite{applphysa,morgenstern09,comstock07,tegeder07}. 

In this situation first-principles calculations stand to offer a unique perspective on the geometric and electronic structure of these systems. With such motivation we have recently performed a density-functional theory (DFT) study of the metastable states of azobenzene adsorbed at coinage metal (111) surfaces \cite{mcnellis09}. Within the employed generalized-gradient approximation (GGA) to the DFT exchange-correlation (xc) functional, the essential findings of this study are summarized as follows: At all three substrates the most stable adsorption geometry corresponds to the azo (-N=N-) bridge centered on and aligned with the bridge site of the (111) lattice, cf. Fig. \ref{fig2} below. In the flat trans isomer geometry the two phenyl ($-$C$_{6}$H$_{5}$) moieties lie parallel to the surface, while in the three-dimensional cis isomer geometry they stand upright, pointing away from the surface. In this decomposition into azo-bridge and phenyl-ring moieties, the overall bonding is characterized by a balance between three major effects: The binding energy gained by the formation of a covalent-type bond between azo-bridge and surface, the energetic penalty due to distortion of the gas-phase molecular geometry upon adsorption, and Pauli repulsion between the phenyl-rings and the substrate. In the flat trans adsorption geometry the first and third effect are in conflict in the sense that the formation of covalent-type bonds requires shorter surface distances, where the phenyl-rings already suffer from strong Pauli repulsion. Since this conflict does not arise in the three-dimensional cis geometry, the latter is relatively more stabilized upon adsorption: While the gas phase trans isomer is more stable than cis by some 0.6 eV, this value is lowered upon adsorption and at Cu(111) -- where the azo-bridge to surface bond is strongest -- DFT-GGA even predicts a reordering of the isomer stability, with the cis isomer $\sim$ 0.3 eV lower in energy \cite{mcnellis09}. 

Within this picture of a delicate balance between competing effects, it is clear that these results will be sensitive to the well-known inability \cite{kristyan94} of semi-local GGA xc-functionals to account for a fourth surface bonding contribution, namely dispersive van der Waals (vdW) interactions. For the present system a rough estimate of the importance of this contribution may be obtained from the similarity of the phenyl moiety and a benzene (C$_6$H$_6$) molecule. Comparing the $\sim$\,0.6 eV benzene binding energy derived from temperature programmed desorption (TPD) experiments \cite{xi94,zhou90,syomin01} to the essentially zero value obtained within DFT-GGA \cite{bilic05}, {\em vide infra}, suggests that the computed DFT-GGA binding energy of the trans azobenzene isomer with its two flat-lying phenyl-rings might be underestimated by more than 1\,eV. While this underscores that a real understanding of azobenzene and related organic molecules at metal surfaces can only be obtained from a properly balanced description of all four surface bonding contributions, realizing this in electronic structure calculations for such systems is still a largely unresolved challenge. In fact, an accurate account of dispersion interactions is one of the major issues in contemporary large-scale first-principles modeling, and recent years have seen a plethora of proposed solutions, ranging from high-level wave-function techniques within various embedding schemes \cite{hu07,sharifzadeh08}, via xc-functionals explicitly constructed to include dispersion interactions \cite{rydberg04,dion04}, to the electron correlation resulting from applying the random-phase approximation (RPA) in the context of the adiabatic-connection dissipation-fluctuation theorem \cite{furche01,rohlfing08}. 

Unfortunately, due to the sheer adsorbate-dictated system size none of these approaches are presently computationally tractable for the problem at hand. In this situation, much more modest semi-empirical schemes that correct at least for the long-range vdW interaction represent a viable alternative for a first assessment of how much the missing fourth bonding contribution skews the previously obtained DFT-GGA results. However, while computationally highly efficient, these semi-empirical dispersion correction approaches \cite{wu02,ortmann06,grimme06,jurecka07,tkatchenko09} are also not unproblematic. On the practical side their semi-empirical derivation has given rise to a manifold of suggested schemes that all have the same conceptual structure, but differ in their material-specific parameters. On the fundamental side the mere validity of these schemes for bonding at metal surfaces is uncertain. Apart from the system-specific interest a second objective of our study is therefore to subject the obtained dispersion corrected results to critical scrutiny. For this we analyze the scatter when applying different prevalent semi-empiricial correction schemes to the azobenzene problem and use the adsorption of benzene at the close-packed coinage metal surfaces as a vehicle to isolate the balance between Pauli repulsion and vdW interaction within this methodology. 

The overall structure of the paper is correspondingly as follows: The next theory section provides a detailed description primarily of the different employed semi-empirical correction schemes and how they are integrated into the DFT calculations. Thereafter we present the changes obtained through these dispersion corrections to the previous pure DFT-GGA results for the geometric, energetic and electronic properties of adsorbed azobenzene at Cu(111), Ag(111) and Au(111). This is followed by a critical discussion which includes the detailed results when applying the semi-empirical correction schemes to the simpler problem of benzene at these three surfaces.

\section{Theory}

\subsection{Semi-empirical dispersion correction schemes}

\begin{figure}
\centering
\includegraphics[width=8.2cm]{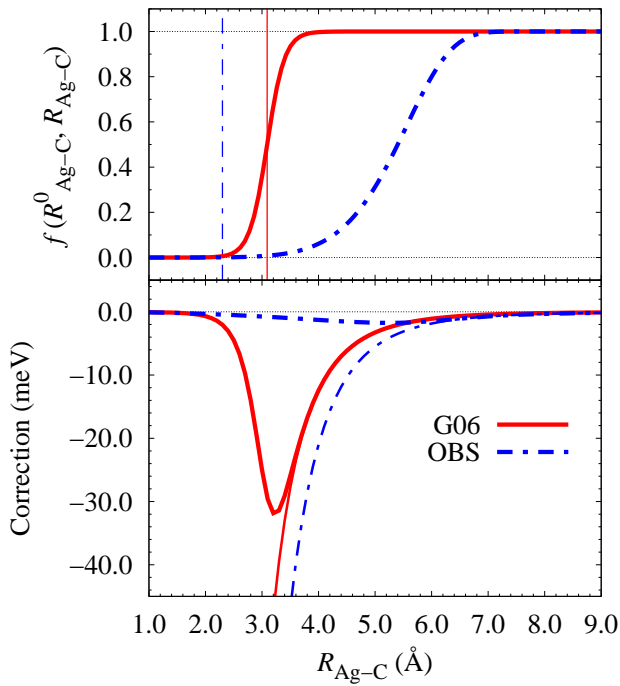}
\caption{(Color online)
Illustration of the dispersion correction between a Ag-C atom pair as a function of their interatomic distance $R_{\rm Ag-C}$. Shown in the lower panel is the $s_6 C_{\rm 6,Ag-C} R_{\rm Ag-C}^{-6}$ term in the OBS (medium dash-dotted line) and G06 (medium solid line) scheme, as well as the finally resulting correction after multiplication with the damping function $f(R^{0}_{\rm Ag-C},R_{\rm Ag-C})$ (separately shown in the upper panel), cf. Eq. (\ref{eq1}): OBS (thick dash-dotted line) and G06 (thick solid line). Note the largely different form of the dispersion correction, which at not too different $C_{6,{\rm Ag-C}}$ coefficients in the two schemes, cf. Table \ref{table0}, is predominantly due to the very shallow damping function employed in the OBS scheme. At distances comparable to the sum of interatomic vdW radii, 2.30\,{\AA} and 3.09\,{\AA} for OBS and G06 respectively (marked as thin vertical lines in the upper panel), this leads to energetic corrections that differ by an order of magnitude. The TS correction potential is qualitatively very similar to that of G06, differing in $s_6 C_{\rm 6}$ and $S_{R}R^{0}$ parameters only.}
\label{fig1}
\end{figure}

In the semi-empirical dispersion correction approach, the missing dispersion contribution to the inter-atomic interaction is approximated by a simple isotropic potential. At long range, this potential is given by the leading $C_{6,ij} \cdot R_{ij}^{-6}$ term of the London series, where $C_{6,ij}$ is a material-specific, so-called dispersion coefficient between any atom pair $i$ and $j$ at distance $R_{ij}$. At short range, the long-range expression is matched to the DFT potential by multiplication with a damping function $f(R^{0}_{ij},R_{ij})$, which reduces the additional dispersion contribution to zero, subject to a cutoff defined by some suitably calculated combination $R^{0}_{ij}$ of the vdW radii of the atom pair, as illustrated in Fig. \ref{fig1}. The dispersion corrected xc-functional is then formed by simply adding the correction potential to the ordinary DFT xc-functional. As $C_{6,ij}$ coefficients are additive\cite{stone96}, the dispersion corrected total energy $E_{\rm tot}$ may therefore be written as
\begin{equation}
\label{eq1}
E_{\rm tot} \!= E_{\rm DFT} + s_6 \sum_{i=1}^{N}\sum_{j>i}^{N} f(S_R R^{0}_{ij},R_{ij}) \; C_{6,ij} \; R_{ij}^{-6} \quad ,
\end{equation}
where $E_{\rm DFT}$ is the standard DFT total energy, and the sums go over all {\em N} atoms in the system. In existing schemes, heteronuclear $R^{0}_{ij}$ and $C_{6,ij}$ coefficients are approximated from semi-empirically determined homonuclear parameters, and differences between DFT xc-functionals in the description of short to medium range dispersion interaction are taken into account by a suitable modification of the correction potential through the parameters $s_6$ or $S_R$ as further detailed below.

While effortless to evaluate computationally, this general expression has obvious weaknesses, of which perhaps the most glaring is the (free-) atom-in-molecule approximation: The substantial variation of properties between effective 'atoms' of the same species in the molecule, e.g. the influence of hybridization states on effective polarizability, is neglected. This may be particularly severe for the metal surface, where the lowering of the effective dielectric constant due to screening \cite{rehr75} should be reflected by reduced dispersion coefficients of atoms in deeper layers in the substrate. The applicability of such semi-empirical dispersion corrections to (transition) metal bonding systems is therefore uncertain. In order to qualify this to some extent for the present azobenzene at coinage metal surface problem, we therefore study the variation of our results between three published dispersion correction schemes that (apart from different semi-empirical material parameters) each have their own approach to this issue:

\begin{center}
\begin{table}
\caption{\label{table0} Calculated $C_{6,ij}$ dispersion coefficients and cutoff radii $R^{0}_{ij}$ between elements $i$ and $j$. As the dispersion coefficients in the TS scheme depend slightly, $\sim \pm 5$\,\% for adsorbed trans azobenzene, on the effective atomic volume we only show averaged values over all atomic constituents of each species. To allow direct comparison between the effective parameters employed by the three schemes, we show the values $s_{6} C_{6,ij}$ and $S_{R} R^{0}_{ij}$, i.e. after multiplication with the xc-functional dependent scaling factors (see text). Note that the G06 scheme does not feature parameters for Au.}
\begin{tabular}{cc|ccc|ccc}
\hline
\hline
	& & \multicolumn{3}{c|}{$s_{6} C_{6,ij}$ [eV$\cdot${\AA}$^6$]} & \multicolumn{3}{c}{$S_{R} R^{0}_{ij}$ [{\AA}]} \\
	&   		           	& H    & C    & N & H    & C    & N \\
\hline
        & OBS                   & 31.9 & 77.7 & 54.0 & 1.81 & 2.21 & 2.19 \\
Au      & G06                   & $-$  & $-$  & $-$  & $-$  & $-$  & $-$  \\
        & $\langle$TS$\rangle$  & 19.7 & 62.0 & 45.9 & 3.48 & 3.80 & 3.71 \\
\hline 
        & OBS                   & 35.0 & 86.1 & 59.2 & 1.90 & 2.30 & 2.28 \\
Ag      & G06                   & 14.5 & 51.1 & 42.8 & 2.64 & 3.09 & 3.04 \\
        & $\langle$TS$\rangle$  & 19.3 & 60.9 & 43.7 & 3.45 & 3.78 & 3.69 \\
\hline
        & OBS                   & 30.6 & 75.0 & 51.6 & 1.75 & 2.15 & 2.13 \\
Cu      & G06                   & 9.5  & 33.8 & 28.3 & 2.56 & 3.01 & 2.96 \\
        & $\langle$TS$\rangle$  & 17.9 & 55.6 & 40.1 & 3.45 & 3.77 & 3.68 \\
\end{tabular}
\end{table}
\end{center}

The first scheme is due to Ortmann, Bechstedt and Schmidt (OBS) \cite{ortmann06}, and employs the London formula \cite{tang69} to calculate the dispersion coefficients from experimentally measured polarizabilities and ionization potentials. The corresponding dispersion coefficients, $C_{6,ij}$ are compiled in Table \ref{table0}, together with those of the other two schemes. The notably shallow, long-ranged damping function in the OBS scheme illustrated in Fig. \ref{fig1} was fitted such that together with $E_{\rm DFT}$ within the GGA of Perdew and Wang (PW91) \cite{pw91} the experimental {\em c}-lattice constant of graphite is reproduced. Demonstrating the effect of such a shallow damping function is the main reason why this scheme is considered here, even though it was not developed for universal transferability and does therefore e.g. not provide fits to other xc-functionals in the original publications\cite{ortmann06} (i.e. $s_6\,=\,S_R \equiv 1$).

The second correction scheme employed is the 2006 revision proposed by Grimme (G06)\cite{grimme06}. It features parameters calculated from first-principles for most of the periodic table (although not including Au), a compared to OBS steeper and shorter ranged Fermi damping function, cf. Fig. \ref{fig1}, and a xc-functional dependence determined by fitting to thermochemical benchmark calculations. This scheme does aim at transferability, and has garnered popularity in the field, with applications including adsorption problems \cite{atodiresei08,atodiresei09} akin to this study. However, for the present work, two G06 design choices may prove detrimental: First, the problem of defining transferable transition element 'atoms', is admittedly \cite{grimme06} crudely solved by taking their parameters as simply the average of those of the preceding noble gas atom and the following group III element. Second, the xc-functional dependence is introduced by (effectively) scaling the dispersion coefficients with a factor $s_{6} \ne 1$ ($S_R \equiv 1$), cf. Eq. (\ref{eq1}), which corrects most for the xc-functional where its influence is smallest, i.e. it shifts the $C_{6,ij} \, R_{ij}^{-6}$ potential also at long range, where it can be expected to be most accurate. 

The third scheme was recently put forth by Tkatchenko and Scheffler (TS) \cite{tkatchenko09}, and represents the state of the art: Exploiting the relationship between polarizability and volume\cite{brink93}, this scheme accounts to some degree for the relative variation in dispersion coefficients of differently bonded atoms. This is achieved by weighting values taken from the high-quality first-principles database of Chu and Dalgarno\cite{chu04}, with atomic volumes derived from Hirshfeld partitioning\cite{hirshfeld77} of the self-consistent electronic density. The TS scheme uses the same damping function as the G06 scheme, but following Jure\v cka {\em et al.} \cite {jurecka07}, instead scales the combined vdW radius by a xc-functional dependent factor $S_{R} \ne 1$ ($s_6 \equiv 1$), thereby correcting where the xc-functional influence is strongest, and leaving asymptotics intact.

\subsection{Density-Functional Theory Calculations}

All DFT calculations were performed with the CASTEP \cite{clark05} code using a plane-wave basis together with ultrasoft pseudopotentials \cite{vanderbilt90} as provided in the default library, and the GGA-PBE functional \cite{perdew96} to treat electronic exchange and correlation. For benzene adsorption the local-density approximation (LDA) in the parameterization by Perdew and Zunger \cite{perdew81} was additionally considered. The computational parameters were exactly those used and detailed already in the preceding study \cite{mcnellis09}, which is why we only briefly recapitulate those aspects of relevance for the dispersion correction schemes here. For azobenzene and benzene we modeled the (111) surfaces as four layer 
slabs in supercells with a lateral periodicity of $(6 \times 3)$ and $(3 \times 3)$ surface unit-cells with 18 and 9 atoms per layer, respectively. Dispersion corrections were implemented in an external module\cite{sedc} such, that a boundary-condition dimensionality $0 \leq D \leq 3$ can be chosen for arbitrarily defined subsets of the supercell atoms. Aiming to describe properties in the low coverage limit, this allows to account for dispersion corrections between the adsorbate and an extended substrate, while simultaneously suppressing spurious long-range dispersion interactions between the adsorbate and its periodic images. Where applicable, parameters optimized for the employed GGA-PBE functional were used in the dispersion correction schemes, i.e. $s_{6} = 0.75$ (G06)\cite{grimme06} and $S_{R} = 0.94$ (TS)\cite{tkatchenko09} was used in Eq. (\ref{eq1}). 

The module was interfaced to a locally modified version of the academic release of CASTEP 4.3 allowing to include the externally calculated dispersion corrections to total energies, forces and stress. Furthermore, it was extended by routines providing the atomic volumes based on a Hirshfeld analysis\cite{hirshfeld77} as required for the TS scheme. The Hirshfeld analysis is based on the (soft) pseudo charge densities for both the superposition of atomic and the self-consistent densities. Hirshfeld charges and volumes are integrated on the (standard) real space grid as provided by the plane wave basis set. We carefully checked the convergence of the volumes with respect to the size of the simulation cell by integrating over appropriate supercells (using properly replicated charge densities in the self-consistent case and applying some special treatment for hydrogen atoms to suppress the numerical noise on the integration grid). For several small organic molecules and the fcc bulk phases of Au, Ag and Cu we have compared our implementation to reference values obtained with the all-electron code FHI-AIMS\cite{aims} which employs numerical atom centered basis functions based on radial grids. For the transition metal (pseudo-) atoms the obtained Hirshfeld volumes differ rather substantially from the reference values. However, the ratios of the two volumes, which are the only required input for the TS scheme, are in very good agreement with the reference numbers in all cases and for all elements in this study. As a side remark, we note that this demonstrates for the first time the applicability (with very modest implementation effort) of the TS scheme to DFT calculations done with a plane wave basis set. Further details on this implementation and the numerical tests will be published elsewhere \cite{hirshfeld_pw}.

As in our preceding study \cite{mcnellis09} the centrally targeted energetic quantities are the adsorption energy $E_{\rm ads}$ and the relative isomer stability $\Delta E$. The former quantity is defined as
\begin{equation}
\label{eq2}
E_{\rm ads} \;=\; \frac{1}{2} \left[ E_{\rm azo@(111)} - E_{(111)} - E_{\rm azo(gas)} \right] \quad ,
\end{equation}
while the latter quantity is defined as
\begin{equation}
\label{eq3}
\Delta E \;=\; \frac{1}{2} \left[ E_{\rm azo@(111)}({\rm cis}) - E_{\rm azo@(111)}({\rm trans}) \right ] \quad.
\end{equation}
Here, $E_{\rm azo@(111)}$ is the total energy of the relaxed azobenzene-surface system, $E_{(111)}$ the total energy of the clean slab, $E_{\rm azo(gas)}$ the total energy of the corresponding relaxed gas-phase isomer (all three computed at the same plane-wave cutoff), and the factor $1/2$ accounts for the fact that adsorption is at both sides of the slab. The adsorption energy of either cis or trans isomer at the surface is thus measured relative to its stability in the gas-phase, and a negative sign indicates that adsorption is exothermic. $\Delta E > 0$, on the other hand, indicates that the trans isomer is more stable at the surface. Systematic convergence tests indicate that these quantities are converged to within $\pm 30$\,meV at the chosen computational settings \cite{mcnellis09}. Defined as differences of surface with and without adsorbate resp. adsorbed trans and cis total energies, $E_{\rm ads}$ and $\Delta E$ are virtually unaffected by the intra-substrate, i.e. metal-metal dispersion corrections. Residual contributions arise from the only slightly different surface atom relaxations upon adsorption and in case of the TS scheme from different Hirshfeld decompositions. Given the anticipated limitations of the semi-empirical approach in describing metal surfaces, we deliberately switch these small contributions off, i.e. the reported dispersion-corrected $E_{\rm ads}$ and $\Delta E$ only consider intra-molecular and molecule-substrate vdW contributions. Here, the intra-molecular dispersion corrections to these quantities arising from the changed molecular geometry in the adsorption structure are in fact minute and are below 15\,meV at all surfaces and for all schemes. For benzene, we additionally calculated the zero-point energy correction to the adsorption energies $E_{\rm ads}^{\rm ZPE}$. The required normal modes of the benzene molecule have been obtained as $\Gamma$-point phonons. In our supercell setup these are computed via finite displacements from equilibrium positions to obtain the nuclear Hessian by numerical differentiation of the resulting forces.

\section{Results}

\subsection{Gas-Phase Azobenzene}

Since the azobenzene adsorption properties are consistently referenced to the gas-phase molecule, it forms a natural starting point for our analysis. As previously reported\cite{mcnellis09}, GGA-PBE yields the gas-phase geometric structure of both isomers in close agreement with B3LYP\cite{becke93} hybrid-functional DFT, many-body perturbation theory calculations and experiment. This result remains unchanged here; none of the correction schemes applied modifies the pure GGA-PBE gas-phase geometries significantly. The gas-phase cis-trans relative energetic stability is also only modestly reduced, from $\Delta E = 0.57$\,eV for pure GGA-PBE, to 0.44\,eV (OBS), 0.47\,eV (G06) and 0.49\,eV (TS) in the three semi-empirical schemes. This result is readily explained by the shorter distance between the phenyl-rings in the three-dimensional cis isomer, which leads to a larger dispersion stabilization in the cis isomer and thereby to a lowering of $\Delta E$. This lowering by $\sim 0.1$\,eV due to the approximate account of dispersive interactions suggests that the near-perfect agreement between GGA-PBE, $\Delta E = 0.57$\,eV, and experiment, $\Delta E = 0.6$\,eV \cite{schulze77}, is due to a fortuitous cancellation between long- and short-range errors in the semi-local functional. This interpretation is corroborated by the $\Delta E = 0.68$\,eV \cite{mcnellis09} obtained with B3LYP, which is intended to improve on the DFT-GGA at short range. As the latter overshoots by $\sim 0.1$\,eV, one could suspect that dispersion corrected B3LYP would then restore the agreement to experiment.

\subsection{Adsorbate geometric structure}

\begin{figure}
\centering
\includegraphics[width=0.4\textwidth]{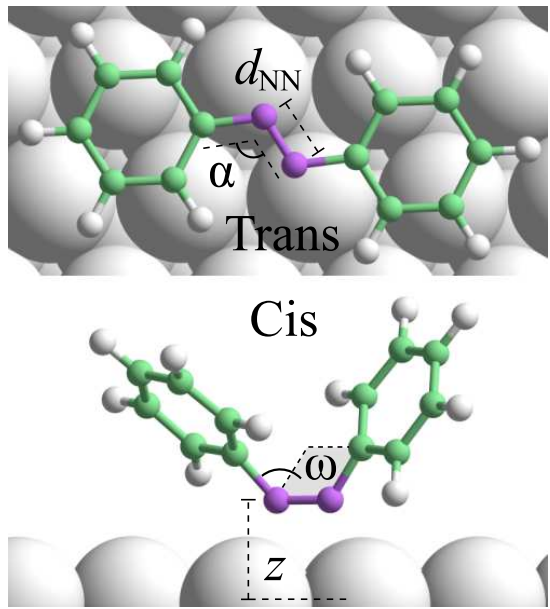}
\caption{(Color online) Top panel: Top view of the azobenzene trans isomer at the preferred DFT-GGA adsorption site. Bottom panel: Side view of the corresponding cis isomer adsorption geometry. Additionally shown in both panels are definitions of analyzed geometry parameters {\em z}, $d_{NN}$, $\omega$, and $\alpha$ (see text).}
\label{fig2}
\end{figure}

Turning to azobenzene adsorbed at the coinage metal surfaces, we first note that due to the long-range nature of the dispersive interactions the energetic correction provided by the semi-empirical schemes only shows a small corrugation with respect to lateral movements of the adsorbed molecule across the surface, leaving the optimal adsorption geometry largely determined by the DFT energetics. Correspondingly, full geometry optimizations of all high-symmetry adsorption geometries as defined in Fig. 1 of our previous publication\cite{mcnellis09}, at all surfaces and using all three dispersion correction schemes, show a clear energetic preference for the previously determined most stable adsorption site at GGA-PBE level of theory. In this geometry, the central azo-bridge is aligned in a 1:1 N-metal coordination as shown in Fig. \ref{fig2}. The resulting adsorption geometry is suitably described by four characteristic parameters also shown in Fig. \ref{fig2}, namely the vertical distance from the top surface layer to the azo-bridge plane {\em z}, the NN bond length $d_{\rm NN}$, the CNNC dihedral angle $\omega$, and the CNN bond angle $\alpha$. At DFT-GGA level the latter angle $\alpha$ remained unchanged from its gas-phase value, 115$^\circ$ and 124$^\circ$ for trans and cis azobenzene respectively, for both phenyl-rings at all three substrates. All three semi-empirical correction schemes similarly leave this angle unaffected, which is why Table \ref{table1} concentrates on the obtained results for the three remaining structural parameters.

\begin{center}
\begin{table}
\caption{\label{table1} Azobenzene structural parameters as defined in Fig. \ref{fig2} and as obtained using GGA-PBE, as well as the three semi-empirical correction schemes due to Ortmann, Bechstedt and Schmidt (OBS), Grimme (G06) and Tkatchenko and Scheffler (TS). None of these schemes affects the gas-phase geometric parameters, which is why only the PBE values are quoted here. Note that the G06 scheme does not feature parameters for Au.}
\begin{tabular}{cc|ccc|ccc}
\hline
\hline
	& & \multicolumn{3}{c|}{Trans} & \multicolumn{3}{c}{Cis} \\
& & $z$ & $d_{\rm NN}$ & $\omega$ & $z$ & $d_{\rm NN}$ & $\omega$ \\
    &    & ({\AA}) & ({\AA}) & (deg) & ({\AA}) & ({\AA}) & (deg) \\
\hline
Gas-phase & PBE & $-$  & 1.30 & 0  & $-$  & 1.28 & 12 \\
\hline
	      & PBE & 3.50 & 1.30 & 0  & 2.31 & 1.29 & 18 \\
Au(111)	& OBS & 3.48 & 1.30 & 0  & 2.24 & 1.30 & 19 \\
        & G06 & $-$  & $-$  & $-$& $-$  & $-$  & $-$ \\
	      & TS  & 3.28 & 1.30 & 0  & 2.23 & 1.30 & 18 \\
\hline 
        & PBE & 3.64 & 1.30 & 0  & 2.27 & 1.32 & 23 \\
Ag(111)	& OBS & 3.60 & 1.30 & 0  & 2.20 & 1.32 & 25 \\
        & G06 & 2.75 & 1.33 & 0  & 2.14 & 1.32 & 25 \\
        & TS  & 2.98 & 1.31 & 1  & 2.16 & 1.32 & 25 \\	
\hline
        & PBE & 1.98 & 1.40 & 39 & 1.93 & 1.35 & 33 \\
Cu(111)	& OBS & 1.97 & 1.40 & 38 & 1.91 & 1.35 & 33 \\
        & G06 & 2.05 & 1.40 & 11 & 1.89 & 1.35 & 35 \\
        & TS  & 2.05 & 1.40 & 13 & 1.89 & 1.35 & 34 \\
\end{tabular}
\end{table}
\end{center}

At DFT-GGA level gas-phase molecular geometries at Au(111) are largely unmodified by adsorption, a fact which dispersion corrections do not change: OBS does not affect the trans isomer at all, and the geometry correction in the other cases is essentially limited to a small rigid shift of the adsorbate towards the substrate. For the trans isomer this means that the dihedral angle $\omega$ remains at its gas-phase value of zero, i.e. the molecule stays planar. At Ag(111), the picture is somewhat more varied: Here, the stronger azo-bridge$-$surface interaction activates and elongates the NN bond in the cis isomer, a result unchanged by the dispersion corrections despite a small downward shift similar in magnitude to the changes of $z$ observed at Au(111). For the trans isomer, correction effects range from none in the OBS scheme, to a dramatic reduction of {\em z} with concomitant elongation of $d_{\rm NN}$ in the G06 and TS schemes. Nevertheless, the trans isomer at Ag(111) again remains planar in all cases. Finally, at Cu(111) the role of dispersion corrections is decidedly different from that at Au(111) and Ag(111): With GGA-PBE already yielding a strong azo-bridge to surface bond, which significantly elongates the NN bond and pins the bridge at a short vertical distance to the surface, the cis isomer {\em z}-shift induced by the dispersion schemes is smaller than at the other two substrates. Since this azo-bridge bond also determines the surface distance in the trans isomer and thereby puts the phenyl-rings well inside the range of surface Pauli repulsion, the intra-molecular distortion energy succumbs and as indicated by the large value for $\omega$, the phenyl-rings are bent out of the molecular plane in GGA-PBE. Again, the OBS scheme is too weak to influence this result. However, the G06 and TS schemes are not, and bend the phenyl-rings back towards the surface, yielding an $\omega$ of some ten degrees, considerably closer to that of the gas-phase geometry (0$^\circ$) than that of GGA-PBE ($\sim$ 40$^\circ$). In fact, this restoring force, self-consistent with the ring repulsion and molecular distortion energy, is sufficient to somewhat offset the effect of the azo-bridge$-$surface bond, which appears as a slightly increased surface distance {\em z}. 

\begin{figure}
\centering
\includegraphics[width=7.6cm]{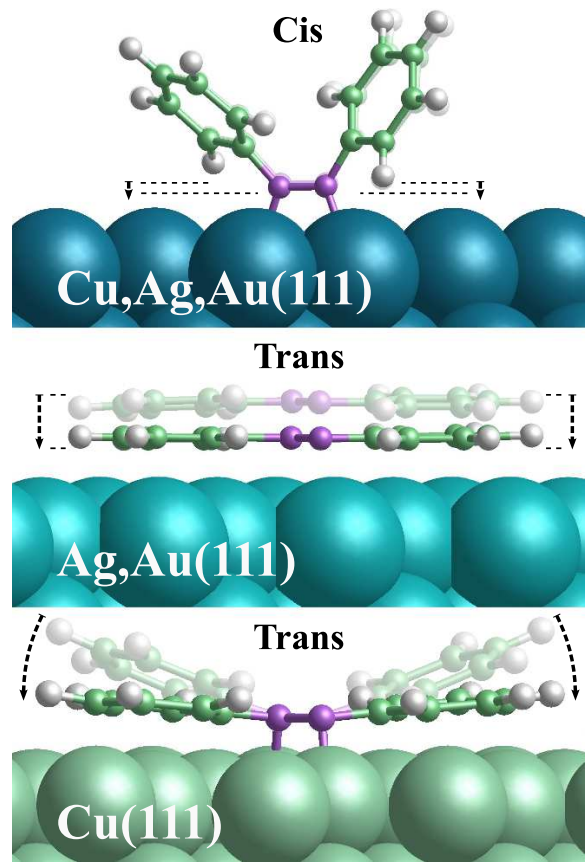}
\caption{(Color online) Schematic illustration of the geometry changes induced by the G06 and TS correction schemes. Shown are sideviews with the GGA-PBE geometries shaded in the background as reference. The OBS scheme does not have a significant effect on the GGA-PBE geometries.}
\label{fig3}
\end{figure}

As schematically summarized in Fig. \ref{fig3} the OBS scheme correction is thus overall too weak to significantly modify molecular geometries. Given its shallow, long-ranged damping function, this result is readily understood: As apparent from Fig. \ref{fig1} the less defined minimum in the correction potential turns the substrate dispersion interactions into a smooth background potential, with small gradient contributions. On the other hand, the G06 and TS schemes use similar, deeper damping functions with gradient corrections strong enough to make adsorbate geometries in principle dependent on the local substrate geometry. However, in the cis isomers the phenyl-rings sit largely outside the large-gradient range of this G06/TS damping function. The geometry corrections are therefore small and practically identical for the two schemes. For the planar trans azobenzene this situation is different and the phenyl-rings do fall inside the large-gradient range. At Au(111) and Ag(111), where the weak azo-bridge$-$surface bond does not fix the molecular height, the resulting attractive dispersion interaction correspondingly pulls down the entire molecule. At Cu(111), the strong azo-bridge$-$surface bond prevents this, and the dispersive attraction only flattens the molecule by bringing the phenyl-rings closer to the surface. 

\subsection{Electronic structure and adsorbate energetics}

\begin{center}
\begin{table}
\caption{\label{table2} Adsorption energies $E_{\rm ads}$ and work functions $\Phi$ for the azobenzene cis (C) and trans (T) isomers at the three coinage metal surfaces, as well as the relative C-T energetic stability $\Delta E$. Note that the G06 scheme does not feature parameters for Au. All numbers in eV.}
\begin{tabular}{cc|cc|c|cc}
\hline
\hline
  & & \multicolumn{2}{c|}{$E_{ads}$} & $\Delta E$ & \multicolumn{2}{c}{$\Phi$}  \\
          &     &   T & C   & C-T   & T   & C  \\
\hline
          & PBE & $-$ & $-$ &  0.57 & $-$ & $-$ \\
Gas-Phase & OBS & $-$ & $-$ &  0.44 & $-$ & $-$ \\
          & G06 & $-$ & $-$ &  0.47 & $-$ & $-$ \\
          & TS  & $-$ & $-$ &  0.49 & $-$ & $-$ \\
\hline
	    & PBE & -0.12 & -0.27 & 0.42 & 4.82 & 3.89  \\
Au(111) & OBS & -1.05 & -0.99 & 0.50 & 4.82 & 3.89  \\
        & G06 & $-$   & $-$   & $-$  & $-$  & $-$   \\
	    & TS  & -1.67 & -1.23 & 0.93 & 4.70 & 3.89  \\
\hline 
        & PBE & -0.11 & -0.42 & 0.26 & 4.20 & 3.68  \\
Ag(111)	& OBS & -1.05 & -1.18 & 0.31 & 4.18 & 3.72  \\
        & G06 & -2.20 & -1.58 & 1.09 & 4.05 & 3.73  \\
        & TS  & -1.76 & -1.41 & 0.85 & 4.05 & 3.73  \\	
\hline
        & PBE & -0.27 & -1.08 & -0.24 & 4.13 & 3.87 \\
Cu(111)	& OBS & -1.97 & -2.25 &  0.16 & 4.12 & 3.86 \\
        & G06 & -3.00 & -2.30 &  1.18 & 3.81 & 3.87 \\
        & TS  & -2.81 & -2.46 &  0.85 & 3.81 & 3.88 \\
\end{tabular}
\end{table}
\end{center}

Despite these partly severe changes of the molecular adsorption geometry, it is interesting to note that central electronic structure quantities like the work function $\Phi$ or projected frontier orbital positions are not much affected. Compared to corresponding GGA-PBE results detailed in our preceding study \cite{mcnellis09} computed work functions in dispersion corrected geometries differ e.g. between $\sim 0.1-0.3$\,eV, while the entire qualitative rationalization in terms of hybridization and charge rearrangement remains untouched. For completeness we therefore only present the quantitative work function values in Table \ref{table2}, but refer to the preceding publication for further details on the surface electronic structure \cite{mcnellis09}. In contrast, and not surprisingly, the adsorption energetics also summarized in Table \ref{table2} are substantially affected by the dispersion correction schemes. The general pattern of these correction effects follows that observed for the adsorption geometries: At Au(111), the pure GGA-PBE adsorption energies are corrected by approximately 1\,eV, with a comparable correction for both isomers with OBS, and the trans isomer correction some 30\% higher in the TS scheme. Consequentially, the GGA-PBE relative isomer stability $\Delta E$ is practically unaltered in the OBS scheme, but is doubled in the TS scheme. The Ag(111) results are qualitatively as well as quantitatively comparable to those at Au(111): Whereas OBS does not differentiate significantly between isomers, G06 and TS do, binding the trans isomer stronger and correspondingly increasing $\Delta E$. Also at Cu(111) the dispersion corrections qualitatively resemble those at the previous two substrates, but are about 1\,eV larger in magnitude. They are therewith in all three schemes so large that they restore the energetic order of the two isomers back to that in the gas-phase, i.e. the inversion with a more stable cis isomer obtained with GGA-PBE does not prevail. 

Apart from dramatically increasing the adsorption energies, the overall prediction of the three semi-empirical schemes compared to the pure GGA-PBE numbers is thus a larger stabilization of the trans isomer upon adsorption. Within the G06 and TS scheme this dispersion corrected over-stabilization of the trans isomer is in fact so large, that the relative cis-trans energetic stability $\Delta E$ at all three substrates is larger than the corresponding gas-phase value. Despite this qualitative agreement, the corrected energetics provided by the three schemes shows quite pronounced quantitative differences. This holds in particular for the OBS scheme, which in the worst cases yields adsorption energies that differ by almost 1\,eV from those of the G06 or TS scheme. The latter two schemes, on the other hand, are sometimes even more consistent in their results than could be expected from the variation of their corresponding scaled $s_6 \, C_{6,ij}$ coefficients in Table \ref{table0}. Considering that the variation to the OBS coefficients is of similar magnitude, we conclude that the decisive factor for the large difference in the obtained energetics from OBS vs. the G06/TS schemes is less the scatter in the dispersion coefficients, but the differences in the damping function.

\section{Discussion}

In light of the preceding first-principles GGA-PBE results on the nature of the azobenzene$-$surface bond \cite{mcnellis09} the intuitive effects one would expect from an additional attractive dispersion interaction on the adsorption geometry and energetics are easily summarized: Adsorption energies should become more exothermic, somewhat more for the trans isomer with its phenyl-rings at closer distance to the surface. The additional interaction should pull the molecule further down, or, where this is prevented by strong covalent azo-bridge$-$surface bonds, should at least tend to flatten the molecular geometry by driving the Pauli-repelled phenyl-rings to smaller $z$-heights. Per construction, this is exactly what the three semi-empirical schemes do when applied to the azobenzene at coinage metal surface problem, and in this respect the results appear plausible. However, much more conspicuous than the qualitative trend is the sheer magnitude predicted. This is most pronounced for the adsorption energies, which in some cases are increased by more than 2\,eV. Despite the intended nature as a semi-empirical 'correction', the dispersion schemes thereby actually provide a contribution to the final adsorption energy that is up to an order of magnitude larger than the original first-principles result. 

\begin{figure}
\centering
\includegraphics[width=7.8cm]{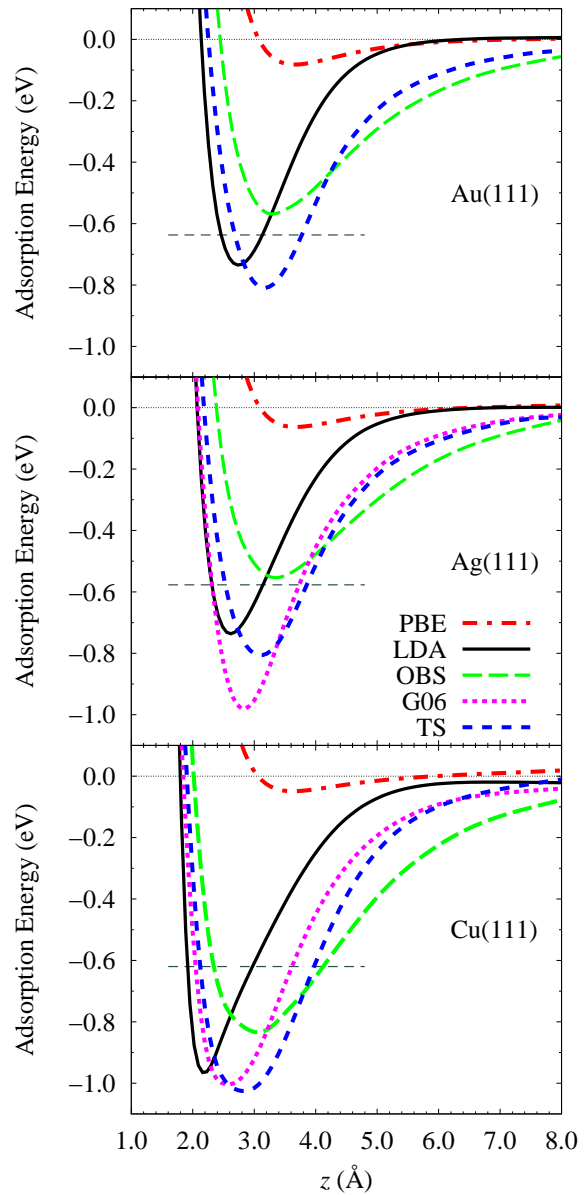}
\caption{(Color online) Binding energy curves of benzene, as a function of the center-of-mass surface distance {\em z}, at all three substrates. Shown is data for LDA and GGA-PBE, as well as the three semi-empirical dispersion correction schemes evaluated at {\em z}-intervals of 0.25\,{\AA}, to which the curves have been fitted as a guide to the eye. For each substrate, the experimentally measured TPD desorption energy is marked as a grey dashed horizontal line (see text).}
\label{fig4}
\end{figure}

Such results can not be uncritically accepted, in particular when recalling the initially summarized fundamental doubts on the applicability of these correction schemes to bonding at metal surfaces at all. In the absence of detailed structural and energetic data for the azobenzene problem either from experiment or high-level theory and as the phenyl-rings play the main part in the trends described, we exploit the similarity of the phenyl-ring moiety ($-$C$_6$H$_5$) to the benzene molecule (C$_6$H$_6$) to get at least some guidance on the order of magnitude for the vdW interactions. For the latter molecule TPD experiments indicate an adsorption energy of $\sim 0.6$\,eV at all three substrates, albeit with rather large error bars\cite{syomin01,zhou90,xi94}. Obtaining again only a minor lateral corrugation of the potential energy surface, we focus here on the so-called 'HCP-B' benzene adsorption geometry \cite{bilic05}, in which the molecule is parallel to the surface \cite{weiss98}, centered over the hcp three-fold hollow site, and rotated such that three carbon atoms are maximally coordinated to the three-fold hollow metal atoms. 

For this geometry, Fig. \ref{fig4} shows the computed binding energy curves for DFT LDA and GGA-PBE, as well as for the three semi-empirical correction schemes. In these calculations only the vertical $z$-height of the molecule was constrained, while all other degrees of freedom were fully relaxed. As in the azobenzene case, the semi-empirical schemes yield sizable corrections at all three substrates that increase the essentially zero GGA-PBE adsorption energies to roughly the spurious LDA value. Also, the trend predicted by the three dispersion schemes is very similar, with OBS adsorption energies consistently lower than the G06 and TS ones. In fact, compared to the data for the more similar trans azobenzene isomer in Table \ref{table1}, the correspondence is almost quantitative: For instance, at Au(111) the adsorption energy correction with respect to GGA-PBE induced by the OBS and TS schemes is 0.46\,eV and 0.72\,eV respectively for benzene, while it is 0.93\,eV and 1.55\,eV for azobenzene with its two phenyl-rings.

\begin{center}
\begin{table}
\caption{\label{table3} Results for benzene at the various substrates, including the experimentally measured adsorption energy $E_{\rm ads}^{\rm exp}$, the computed adsorption energy without $E_{\rm ads}$ and with zero-point energy correction $E_{\rm ads}^{\rm ZPE}$, as well as the optimum molecular height {\em z}.}
\begin{tabular}{c|c|c|ccc}
\hline
\hline
  & $E_{\rm ads}^{\rm exp}$ &   & $E_{\rm ads}$ & $E_{\rm ads}^{\rm ZPE}$ & {\em z} \\
	& (eV) &  & (eV) & (eV) & ({\AA}) \\
\hline
        &                     & PBE & -0.08 & -0.03 & 3.77 \\
        &                     & LDA & -0.73 & -0.68 & 2.75 \\
Au(111) & -0.64\cite{syomin01}& OBS & -0.54 & -0.50 & 3.48 \\
        &                     & G06 & $-$   & $-$   & $-$  \\
	    &                     & TS  & -0.80 & -0.77 & 3.25 \\
\hline 
        &                     & PBE & -0.06 &  0.00 & 3.75 \\
        &                     & LDA & -0.72 & -0.67 & 2.66 \\
Ag(111)	& -0.58\cite{zhou90}  & OBS & -0.55 & -0.48 & 3.38 \\
        &                     & G06 & -0.97 & -0.92 & 2.80 \\
        &                     & TS  & -0.80 & -0.75 & 3.01 \\	
\hline
        &                     & PBE & -0.06 & -0.01 & 3.48 \\
        &                     & LDA & -0.96 & -0.90 & 2.18 \\
Cu(111)	& -0.62\cite{xi94}    & OBS & -0.82 & -0.75 & 3.03 \\
        &                     & G06 & -1.00 & -0.92 & 2.63 \\
        &                     & TS  & -1.05 & -0.98 & 2.75 \\
\end{tabular}
\end{table}
\end{center}

This corroborates benzene adsorption as a suitable reference system and it is therefore meaningful to assess the performance of the three semi-empirical dispersion schemes with respect to the benchmark provided by the TPD experiments. For a more quantitative comparison Table \ref{table3} compiles the corresponding data of the minima of the binding energy curves and additionally accounts for zero-point energy shifts. As obvious from these results, all three schemes certainly provide an improvement compared to the GGA-PBE data, with the OBS scheme even falling close to the experimentally measured value at all three surfaces. The other two schemes exhibit instead a tendency to overbind which gets most pronounced at Cu(111). At this surface this in fact amounts to an overcorrection by $\sim 0.3-0.4$\,eV particularly by these two presumably more refined correction schemes. Within the simple analogy considering only the number of phenyl-rings, this would suggest that the corresponding trans azobenzene at Cu(111) adsorption energies in Table \ref{table1} could contain sizable errors of the order of $\sim 0.6-0.9$\,eV.

This trend of increased overbinding for benzene at Au(111) to Cu(111) goes hand in hand with a systematically smaller height $z$ of the molecule above the surface, cf. Table \ref{table3}. This brings the corresponding molecule$-$metal atom distances entering Eq. (\ref{eq1}) closer and closer to the sum of their vdW radii and therewith into the distance range most affected by the damping function, cf. Fig. \ref{fig1}. In order to assess the implications also for the azobenzene problem, we note that at these intermediate distances this dependence on $f(R^{0}_{ij},R_{ij})$ extends not only to the actual functional form employed, but also sensitively to the cut-off radii\cite{wu02,neumann05}. Thus, while e.g. the metal-carbon correction energy at long range varies exclusively with the rather well motivated $C_{6,ij}$ coefficients (listed in Table \ref{table0}), it may easily in- or decrease by a factor two with just a $\pm$ 10\% variation in $R^{0}_{ij}$ when the metal-carbon distance is comparable to $R^{0}_{ij}$. We would therefore again attribute most of the observed scatter among the three semi-empirical schemes to the specifics of the employed damping function.

In this respect it is discomforting to realize that this influential part of the semi-empirical correction expression, Eq. (\ref{eq1}), is the one least motivated from first-principles. Introduced to merely prevent double counting of those dispersion interaction contributions that are already contained in the employed local or semi-local DFT xc functional at short and intermediate range, there is no guidance on the exact form, nor even existence, of an analytical function that would achieve this. The difference in our OBS versus the G06 and TS results shows that a certain steepness of the damping function is required to achieve a significant dispersion-induced change of the adsorption geometry. On the other hand, we verified that the dependence on the onset of the damping function, i.e. the cutoff-radii, is intuitive. Larger cutoff-radii reduce the dispersion correction, which means for the systems studied here that the adsorption energy is reduced and the adsorption geometry comes closer to the original GGA-PBE one. However, without knowing the exact structure and binding energy, it is impossible to conclude in which direction the damping function employed by the different schemes would need to be modified, even if other uncertainties of the approach like the neglect of metallic screening could be excluded as the source of the error.

Only corresponding detailed data from experiment or high-level theory can therefore provide the final answer as to the performance of the semi-empirical correction schemes for the azobenzene at coinage metal surface problem. In the absence of such data we cautiously conclude from the comparison to the similar benzene adsorption problem that the three schemes employed in this work seem to provide an account of dispersive vdW interactions that is in the right ballpark, and could even be semi-quantitative. The OBS scheme is likely mostly limited by its shallow damping function that yields too small gradients to significantly modify the adsorption geometries. G06 and TS on the other hand are likely to overbind, with a maximum estimated error of $\sim 0.6-0.9$\,eV in the adsorption energies.

\section{Summary}

The problem of azobenzene adsorption at coinage metals combines a large molecule, heavy transition metal surface chemistry and a multi-facetted bonding mechanism, in which dispersive vdW interactions play a crucial role. This combination forms
a tremendous challenge for contemporary first-principles modeling, with DFT with semi-local xc-functionals still setting the standard. In order to assess the effect of the insufficient description of dispersive interactions at this level we employed three different semi-empirical correction schemes that account for such interactions at least on the level of the leading long-range $R^{-6}$ term. The low computational cost of these schemes makes them an attractive solution for this problem, and the overall trend obtained by applying them to the azobenzene problem at Cu(111), Ag(111) and Au(111) is consistent with the anticipated effects of an additional attractive interaction: Compared to pure GGA-PBE adsorption energies become more exothermic, somewhat more for the trans isomer with its phenyl-rings at closer distance to the surface and thereby reinforcing the higher gas-phase trans isomer stability. The additional interaction pulls the molecule further towards the surface, or where this is prevented by strong covalent azo-bridge$-$surface bonds, at least flattens the molecular geometry by driving the Pauli-repelled phenyl-rings to smaller heights.

Much more problematic than this overall trend is the sheer magnitude of the effects predicted. Constructed as a semi-empirical 'correction', the dispersion schemes thereby actually provide a contribution to the final adsorption energy that is in some cases up to an order of magnitude larger than the original first-principles result. In addition the three conceptually similarly rooted schemes exhibit a discomforting scatter in their results that amounts up to 1\,eV in the adsorption energies and up to 1\,{\AA} in central geometry parameters. An analysis of the congeneric adsorption of benzene at the three coinage metal surfaces suggests that the three schemes employed in this work seem to provide an account of dispersive vdW interactions in the azobenzene adsorption that is in the right ballpark, with the more refined G06 and TS schemes likely to overbind. The ultimate answer as to the performance can only be provided by comparison to detailed structural and energetic data from high-level theory or experiment, which could then critically stimulate further development and improvement of these schemes.

\section{Acknowledgements}

Funding by the Deutsche Forschungsgemeinschaft through Sfb 658 - Elementary Processes in Molecular Switches at Surfaces - is gratefully acknowledged. We thank the DEISA Consortium (co-funded by the EU, FP6 projects 508830/031513) for support within the DEISA Extreme Computing Initiative (www.deisa.org), and Dr. A. Tkatchenko for useful discussions.


\begin{thebibliography}{99}
\bibitem{sedc}
This full-featured FORTRAN 90 module is purposely written for simple interfacing to periodic boundary condition or cluster DFT codes. It is licensed under the Lesser GNU Public License (LGPL), and freely available from the authors upon request.
\end{thebibliography}
\end{document}